\RequirePackage{fix-cm}
\documentclass[namedreferences,hyperref,optionalrh]{spr-sola}

\usepackage{booktabs}
\usepackage{graphicx}        
\usepackage{amssymb}        
\usepackage{amsmath} 

\makeatletter
\AtBeginDocument{%
  \def\@tabular{\leavevmode\hbox\bgroup$%
    \let\@acol\@tabacol
    \let\@classz\@tabclassz
    \let\@classiv\@tabclassiv
    \let\\\@tabularcr
    \@tabarray}%
}
\makeatother

\renewcommand{\vec}[1]{{\mathbfit #1}}

\newcommand{\vol}{{\mathcal V}}
\newcommand{\bndry}{{\mathcal S}}

\chardef\us=`\_

\newcommand{\lci}{\ensuremath{\lambda_{\rm ci}}}
\newcommand{\tci}{\ensuremath{\tau_{\rm ci}}}
\newcommand{\arcsec}{\ensuremath{^{\prime\prime}}}
\newcommand{\farcs}{.\!\!^{\prime\prime}}

\newcommand{\dkist}{DKIST}
\newcommand{\muramcode}{MURaM}
\newcommand{\halpha}{H$\alpha$}
\newcommand{\caii}{Ca\,\textsc{ii}}

\newcommand{\vbi}{VBI}
\newcommand{\mfbd}{MFBD}
\newcommand{\flct}{FLCT}
\newcommand{\lct}{LCT}
\newcommand{\ccw}{CCW}
\newcommand{\cw}{CW}
\newcommand{\sdo}{SDO}
\newcommand{\hmi}{HMI}

\begin{document}

\begin{frontmatter}
\title{\dkist{} unveils MHD-predicted sub-50 km photospheric vortices}

\author[addressref=aff1,corref,email={svargasd@unal.edu.co}]
{\inits{S.}\fnm{S.}\,\snm{Vargas\,Domínguez}}

\author[addressref=aff1]
{\inits{O.A.}\fnm{O.\,A.}\,\snm{Calvo\,Rebellon}}

\author[addressref=aff2]
{\inits{J.S.}\fnm{J.\,S.}\,\snm{Castellanos\,Durán}}

\author[addressref=aff2]
{\inits{M.}\fnm{M.}\,\snm{van\,Noort}}

\author[addressref=aff4]
{\inits{F.}\fnm{F.}\,\snm{W\"oger}}

\author[addressref=aff1]
{\inits{D.A.}\fnm{D.\,A.}\,\snm{Rodríguez\,Torres}}

\author[addressref=aff3]
{\inits{C.G.}\fnm{C.\,G.}\,\snm{Bernal}}

\address[id=aff1]
{Universidad Nacional de Colombia, Observatorio Astronómico Nacional, Bogotá, Colombia}

\address[id=aff2]
{Max-Planck-Institut f{\"u}r Sonnensystemforschung, 37077 G{\"o}ttingen, Germany}

\address[id=aff3]
{Universidad Nacional de Colombia, Departamento de Física, Bogotá, Colombia}

\address[id=aff4]
{National Solar Observatory, 3665 Discovery Drive, Boulder, CO 80303, USA}

\runningauthor{Vargas Domínguez et al.}
\runningtitle{\textit{Solar Physics} Example Article}

\begin{abstract}
Solar vortices are rotating plasma structures that play a key role in the
  transport of energy and magnetic helicity from the photosphere to the
  chromosphere and corona. While numerical simulations predict the existence of
  small-scale photospheric vortices with diameters of $50$--$100$\,km, no direct
  observational evidence has been reported to date, primarily due to the limited
  spatial resolution of previous instrumentation.
Here, we analyze high-resolution observations taken by the 4-meter Daniel K. Inouye Solar Telescope (\dkist) at 416\,nm with a 12.3 km spatial resolution.  Using Fourier local correlation
  tracking and the swirling strength criterion ($\lci$),  we identify
  201 vortex detections over a sequence of 143 reconstructed frames
  (${\approx}3.2$\,min) in a field-of-view of $4.92 \times 4.13$\,Mm,
  covering a photospheric plage region with pore-like magnetic concentrations.
  The detected vortices have a median equivalent diameter of $38.6$\,km
  (mean $39.5 \pm 3.3$\,km), with virtually all detections below 50\,km, therefore in the
  sub-granular regime predicted by MHD simulations, but never previously
  observed. The median swirling period $\tci = 37.3$\,s satisfies the
  $\tci < 100$\,s criterion established by \muramcode{} simulations. No statistically significant preference for prograde
  (53.7\%) over retrograde (46.3\%)
  rotation is found, consistent with the negligible
  role of the Coriolis force at sub-granular scales.
  These results demonstrate the unique capability of \dkist{} to resolve the
  photospheric convective vortex population predicted by state-of-the-art
  magnetoconvection simulations, opening a new observational window on
  small-scale solar convective dynamics.  
\end{abstract}
\makeatletter
\@ifundefined{@keywords}{%
}{%
}
\makeatother
\end{frontmatter}

\section{Introduction}
     \label{S-Introduction} 
Vortical motions in the solar atmosphere have been identified across a wide range of spatial and temporal scales, from large-scale vortices of several
megameters (e.g., \citet{Su2012}; \citet{WedB2012}) to small photospheric swirls of a few hundred kilometers
\citep{Bonet2008, WedB2009}. These rotating plasma structures are thought to play a fundamental role in the transport of energy and magnetic helicity along flux tubes connecting the photosphere to the upper solar
atmosphere, potentially contributing to chromospheric and coronal heating
\citep{WedB2012, Shelyag2011, Tziotziou2023}.

State-of-the-art magnetoconvection simulations with the \muramcode{} code \\
\citep{Voegler2005} have made specific predictions about photospheric vortices
at the smallest scales. \citet{Moll2011} demonstrated that the intergranular
network harbors a rich population of vortex tubes with diameters of
$50$--$100$\,km and lifetimes of tens to hundreds of seconds. \citet{Yadav2021},
simulating a unipolar plage region with a mean field of 200\,G at 10\,km grid
resolution, found that sub-granular vortices are denser and hotter than their
surroundings in the chromosphere, and that strong electric current sheets at
their interfaces%
provide an efficient mechanism for chromospheric heating. Those
authors explicitly noted that these structures had never been observed directly
due to the insufficient spatial resolution of existing instrumentation.

Prior observational studies have been fundamentally limited by angular
resolution. \citet{Bonet2008} identified vortices of 0.5--1\arcsec{}
(360--720\,km) with the Swedish Solar Telescope (SST). \citet{Balmaceda2010} and  \citet{VargasDominguez2015} tracked magnetic bright points dragged into photospheric vortex centers using SST G-band and Ca\,\textsc{ii}\,H imagery. \citet{VargasDominguez2011} provided
the first statistical characterization of \lct-derived vortex-like motions in a
quiet-Sun region. The SUNRISE balloon-borne observatory pushed the frontier to
structures of ${\approx}150$\,km at 300\,nm \citep{YellesChaouche2020}, extending
the turbulent power spectrum by more than a factor of two in wavenumber space.
\citet{Kitiashvili2012} showed with NST/BBSO observations at 77\,km resolution
that agreement with simulated kinetic energy spectra requires an effective
resolution of at least 150\,km, still insufficient to resolve the sub-granular
vortex population.

\begin{figure*}
  \centering
  \includegraphics[width=\linewidth]{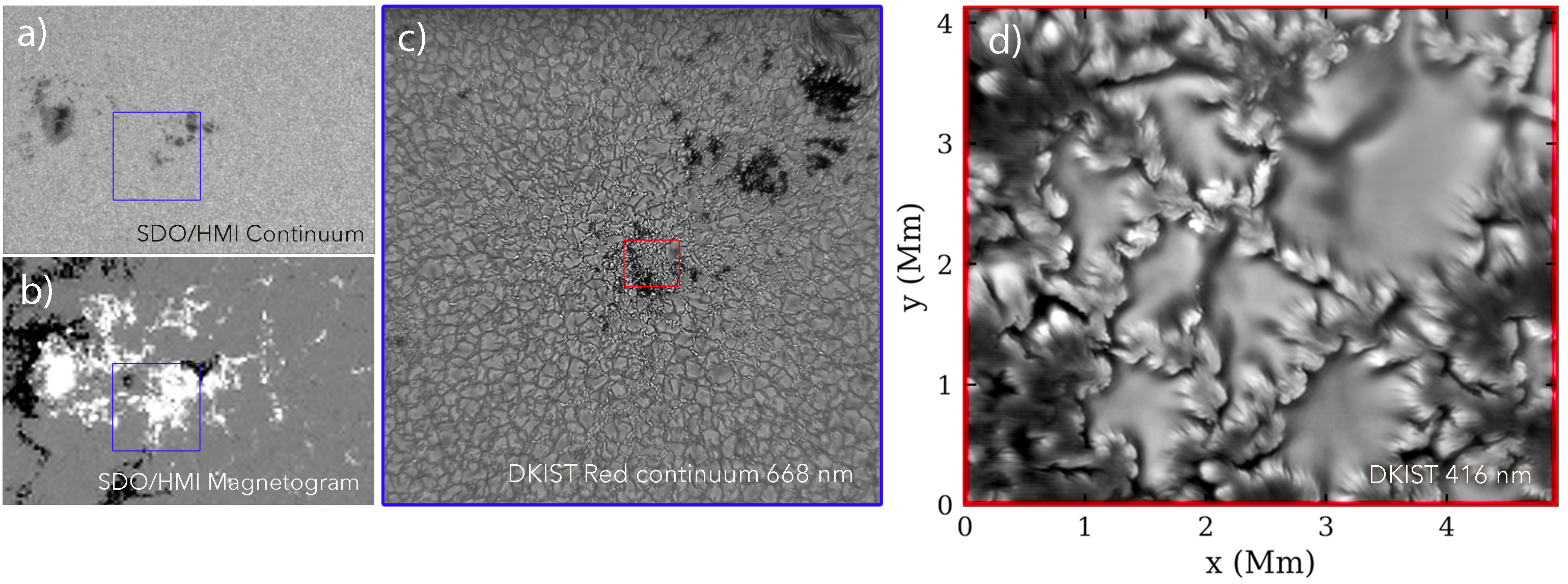}
 \caption{Multi-wavelength context of the observed solar region on 2025 April\,14
at 21:33\,UT, centred near heliographic position
$(\mu = -162\farcs2,\, +168\farcs5)$ in the north-western quadrant.
(a)\,SDO/HMI continuum intensity at 617\,nm showing the large-scale
photospheric structure; the blue box indicates the field of view of the
DKIST/VBI red-continuum image shown in panel\,(c).
(b)\,SDO/HMI line-of-sight magnetogram saturated at $\pm500$\,G;
the blue box is the same as in panel\,(a), marking the DKIST/VBI
field of view; the target region is associated with a small unipolar
flux concentration adjacent to a solar pore, with no evidence of strong
flux emergence during the observing window.
(c)\,DKIST/VBI red-continuum image at 668\,nm showing the
photospheric granulation and the solar pore at high spatial resolution;
the red box indicates the field of view of the DKIST fast-camera sequence
shown in panel\,(d).
(d)\,DKIST fast-camera image at 416\,nm (Ca\,{\sc ii} H wing,
bandpass 0.5\,nm FWHM) used for the vortex detection analysis; the field of
view covers $4.92 \times 4.13$\,Mm after reconstruction with the
multi-frame blind deconvolution (MFBD) technique and removal of the
instrumental padding.
Solar north is up and east is to the left in all panels.}
  \label{fig:context}
\end{figure*}

The Daniel\,K.\ Inouye Solar Telescope \citep[\dkist;][]{Rimmele2020}, with its
4-meter primary mirror, represents a four-fold improvement in aperture
over SUNRISE \citep{Korpi-Lagg2025SoPh..sunrise} and more than an order of magnitude over Hinode/SOT \citep{Ichimoto2008SoPh}. In its
highest-resolution configuration, \\ \dkist{} achieves a plate scale of
${\approx}6$\,km\,pixel$^{-1}$ at visible wavelengths, placing the sub-granular
vortex scales predicted by \muramcode{} within observational reach for the first time; in the same DKIST observing sequence analyzed here,
\citet{Kuridze2026} visually identified prominent vortex-like structures.

In this work, we present the first detection and statistical characterization of
sub-granular vortices in direct photospheric observations using \dkist{} data at
416\,nm. Section\,\ref{sec:obs} describes the observations and context data.
Section\,\ref{sec:methods} presents the detection and tracking pipeline.
Section\,\ref{sec:results} reports the results. Section\,\ref{sec:discussion}
discusses implications. Section\,\ref{sec:conclusions} summarizes our conclusions.

\section{Observations}
\label{sec:obs}


The observations were obtained on 2025 April\,14 starting at 21:33:37\,UT with
\dkist{} at Haleakal{\=a}, Hawaii. The primary dataset consists of high-cadence
intensity images at 416\,nm acquired with a diagnostic fast-camera setup (MPS
FastCam\footnote{\url{dkist.virtualsolar.org/vanNoortfastcam/}}; Jai Spark SP12000CXP4) temporarily installed in front of the Visible
Spectro-Polarimeter (ViSP) spectrograph entry slit, in collaboration with the
Max-Planck-Institut f{\"u}r Sonnensystemforschung (MPS), G{\"o}ttingen. The
bandpass filter had a full width at half maximum (FWHM) of 0.5\,nm centered at
416\,nm (Alluxa ultra-narrow bandpass, OD6). Images were acquired at 740\,Hz. Multi-frame blind deconvolution
(\mfbd; \citealt{Lofdahl2002}) reconstructions were computed from overlapping
sets of 2000 raw frames offset by 1000 frames, yielding a pseudo-cadence of
$\Delta t = 1000/740 \approx 1.35$\,s. The total sequence covers
${\approx}3.2$\,min ($N = 143$ frames). The plate scale is
$\Delta x = 0.00848$\arcsec\,pixel$^{-1}$ (i.e., 6.15\,km\,pixel$^{-1}$ at disk
center).


The observed target is a photospheric plage region with pore-like magnetic flux
concentration at helioprojective coordinates $(-162\arcsec, +168\arcsec)$. The multi-scale observational context
is shown in Fig.\,\ref{fig:context}. The large-scale magnetic environment was characterized with the Helioseismic and
Magnetic Imager \citep[\hmi;][]{Scherrer2012} aboard \sdo. Co-temporal \hmi{}
continuum images and line-of-sight magnetograms ($0.5$\arcsec\,pixel$^{-1}$,
cadence 45\,s) confirm that the \dkist{} field samples a unipolar plage region
of negative polarity, with line-of-sight flux densities of $-50$ to $-400$\,G,
directly comparable to the 200\,G mean-field simulation of \citet{Yadav2021}. High-resolution photospheric context was obtained simultaneously
with the Visible Broadband Imager (\vbi) in the red continuum 668\,nm
(${\approx}50 \times 23$\arcsec). The \vbi{}
red-continuum frames reveal dark pore-like structures of equivalent diameter
1500--3400\,km in the immediate surroundings of the 416\,nm FOV.


The \mfbd{} reconstruction pads the image borders with a constant value of
562.26\,counts. This region was identified by detecting rows and columns with an
intensity standard deviation below 5\,counts; a 10-pixel safety margin was then
applied. Each frame was cropped to rows\,49--849, columns\,50--721 in the raw
FITS array ($671 \times 800$\,pixels) and then
aligned with solar North up and East to the left. The resulting analysis array is
$800 \times 671$\,pixels, corresponding to a physical FOV of
$4.92 \times 4.13$\,Mm (solar X\,$\times$\,Y) = 20.3\,Mm$^2$. Intensity stability
across the full sequence is excellent, with rms $= 2.35$\,counts over a mean of
${\approx}556$\,counts.  The methodology for detecting vortices, which employs Fourier local correlation tracking \citep[\flct;][]{Fisher2008} and the swirling strength criterion \citep[$\lci$;][]{Zhou1999, Haller2005}, in the methods Section~\ref{sec:methods}.

\section{Methods}
\label{sec:methods}

\subsection{Velocity field estimation}
\label{sec:methods:lct}

Horizontal velocity fields were computed from consecutive frame pairs using
Fourier local correlation tracking \citep[\flct;][]{Fisher2008}. Each frame
was normalized to its spatial mean prior to tracking. The algorithm used a
Gaussian apodization window of 20\,pixels (${\approx}123$\,km) and a post-tracking
5-pixel spatial smoothing filter. The displacement field $\boldsymbol{u}=(u_x,u_y)$
was converted to physical velocity via
\begin{equation}
  \boldsymbol{v} = \frac{\Delta x}{\Delta t}\,\boldsymbol{u},
  \label{eq:vel}
\end{equation}
yielding horizontal velocities in the range $-9$ to $+10$\,px\,frame$^{-1}$
(${\approx}-45$ to $+50$\,km\,s$^{-1}$), consistent with published granular flow
velocities in plage regions \citep{VargasDominguez2011}. The time-averaged
horizontal velocity field over the full 3.2-min sequence is shown in
Fig.\,\ref{fig:flowmap}a.


\begin{table}[t]
  \caption{Vortex detection pipeline parameters.}
  \label{tab:pipeline}
  \begin{tabular}{ll}
    \toprule[1.5pt]
    Parameter & Value \\
    \midrule[1.5pt]
    LCT window         & 20\,px (${\approx}123$\,km) \\
    LCT sigma          & 5\,px                      \\ 
    Velocity smoothing & 5\,px                     \\ 
    Plate scale        & 6.15\,km\,px$^{-1}$       \\ 
    Cadence            & 1.35\,s                  \\
    $\lci$ threshold   & 99th percentile  \\
                       & ($= 0.232$\,s$^{-1}$)    \\
    Min.\ $\tci$       & $\approx27$\,s   $= 2\pi/\lci^{\rm thr}$       \\ 
    Min.\ region size  & 28\,px ($r_{\rm eq}\gtrsim18$\,km) \\ 
    Morphol.\ closing  & disk $r = 2$\,px          \\ 
    Max.\ drift        & 60\,km \\ 
    Max.\ radius ratio & $< 2.5$   \\  
    \bottomrule[1.5pt]
  \end{tabular}
\end{table}

\subsection{Vortex detection}
\label{sec:methods:lci}

Vortices were identified using the swirling strength criterion
\citep[$\lci$;][]{Zhou1999, Haller2005}, which detects genuinely rotating flow
independently of local shear. For the velocity gradient tensor
\begin{equation}
  \mathbf{J} =
  \begin{pmatrix}
    \partial v_x/\partial x & \partial v_x/\partial y \\
    \partial v_y/\partial x & \partial v_y/\partial y
  \end{pmatrix},
  \label{eq:Jtensor}
\end{equation}
the discriminant is $\Delta = (\mathrm{tr}\,\mathbf{J})^2 - 4\,\det\mathbf{J}$.
When $\Delta < 0$ the eigenvalues are complex conjugates and
\begin{equation}
  \lci = \tfrac{1}{2}\sqrt{-\Delta}
  \label{eq:lci}
\end{equation}
quantifies the local swirling rate (rad\,s$^{-1}$), with swirling period
$\tci = 2\pi/\lci$. 

We emphasize that $\lambda_{\rm ci}$ (and hence $\tci$) is an instantaneous diagnostic of the local velocity-gradient tensor, computed independently for each frame pair. It quantifies the instantaneous rate of rotation of the flow, analogous to an instantaneous angular velocity, and does not require the underlying rotational pattern to persist for a full period $2\pi/\tci=37.3$\,s. A vortex detected in a single frame with $\tci=37.3$\,s indicates that the local flow is rotating at a rate corresponding to one revolution per 37.3 s at that instant, independently of how long the structure itself survives (see Sect. 4 for further discussion of tracked lifetimes).

The rotation sense (\ccw{} or \cw) follows from the sign
of the vorticity
\begin{equation}
  \omega = \partial v_y/\partial x - \partial v_x/\partial y.
  \label{eq:vorticity}
\end{equation}

The detection threshold is the 99th percentile of the global $\lci$
distribution over all 142 frame pairs and all pixels in the field, giving
$\lci^{\rm thr} = 0.232$\,s$^{-1}$ and a minimum detectable swirling period
$\tci^{\rm min} \approx 27$\,s, well within the $\tci < 100$\,s criterion of
\citet{Yadav2021}.

\subsection{Vortex characterization and tracking}
\label{sec:methods:char}

\begin{figure}
  \centering
  \includegraphics[width=0.97\linewidth]{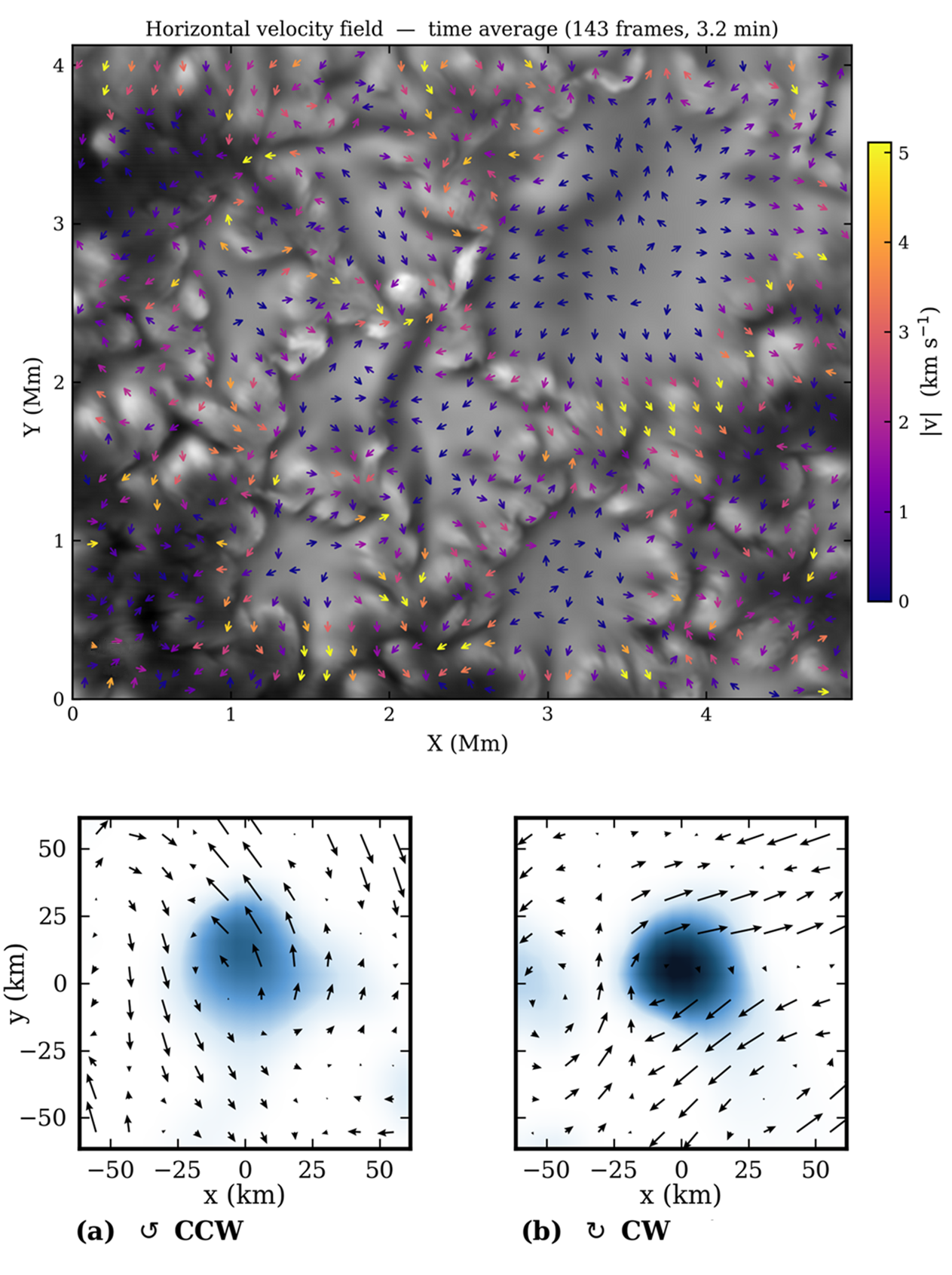}\\[4pt]
 \caption{Horizontal velocity field derived from local correlation tracking (LCT)
applied to the full DKIST 416\,nm sequence (143 frames, cadence 1.35\,s).
\textit{Top:}\,Time-averaged horizontal velocity map over the full FOV
($4.92 \times 4.13$\,Mm). Arrows indicate the direction and relative magnitude
of proper motions, and the background image shows the mean intensity.
The typical flow pattern associated with granular convection cells is clearly
visible, with diverging outflows at granule centres and converging inflows at
intergranular lanes.
\textit{Bottom left} and \textit{bottom right:}\,Close-up views of two
representative detected vortices, one counter-clockwise (CCW) and one clockwise
(CW), respectively. The background image in each panel shows the local
swirling-strength map $\lambda_{\rm ci}$,
and the arrows show the instantaneous horizontal velocity field. The rotational sense of each vortex is clearly reflected in the curling pattern of the velocity arrows.}
  \label{fig:flowmap}
\end{figure}

Connected regions above threshold were extracted after morphological closing
with a 2-pixel disk and removal of objects smaller than 28\,pixels
($r_{\rm eq} \lesssim 18$\,km). For each gion we compute the equivalent radius
$r_{\rm eq}=\sqrt{A/\pi}\,\Delta x$, circularity $C=4\pi A/P^2$, mean
swirling strength $\langle\lci\rangle$, swirling period
$\tci=2\pi/\langle\lci\rangle$, and rotation sense from
$\mathrm{sgn}(\langle\omega\rangle)$. Vortices were tracked across consecutive
frames with a nearest-neighbor algorithm imposing a maximum centroid drift of
60\,km and a maximum equivalent-radius ratio of 2.5. Tracks shorter than two
frames were discarded. Representative examples of individually resolved \ccw{}
and \cw{} vortices are shown in Fig.\,\ref{fig:flowmap}b--c. All pipeline
parameters are summarized in Table\,\ref{tab:pipeline}.

\section{Overview of vortex detection}

A summary of the results of vortex detection is provided in Tab.\,\ref{tab:results}

 \begin{table}[htbp]
   \caption{Summary of vortex detection results.}
   \label{tab:results}
   \begin{tabular}{ll}
     \toprule[1.5pt]
     Quantity & Value \\
     \midrule[1.5pt]
     Observing date         & 2025 April\,14 \\
     Wavelength             & 416\,nm \\
     Plate scale            & 6.15\,km\,px$^{-1}$ \\
     Cadence                & 1.35\,s \\
     Total frames           & 143 \\
     Duration               & 191.9\,s (3.2\,min) \\
     FOV                    & $4.92\times4.13$\,Mm (20.3\,Mm$^2$) \\
     \midrule
     Total detections       & 201 \\
     Mean per frame         & 1.42 \\
     Max.\ per frame        & 7 (frame\,26) \\
     Spatial density        & 0.070\,Mm$^{-2}$ \\
     \midrule
     Median diameter        & 38.6\,km \\
     Mean $\pm$ std diam.   & $(39.5 \pm 3.3)$\,km \\
     Max.\ diameter         & 55.9\,km \\
     Fraction $<50$\,km      & ${>}99\%$ \\
     Fraction $<150$\,km     & 100\% \\
     \midrule
     $\lci^{\rm thr}$       & $0.232$\,s$^{-1}$ \\
     Min.\ $\tci$           & ${\approx}27$\,s \\
     Median $\tci$          & 37.3\,s \\
     \midrule[1.5pt]
     \ccw{}                 & 53.7\% (108 events) \\
     \cw{}                  & 46.3\% (93 events) \\
     Binomial $p$           & 0.29 \\
     \midrule
     Tracked vortices       & 190 \\
     Median lifetime        & 1.35\,s \\
     Maximum lifetime       & 4.05\,s \\
     \bottomrule[1.5pt]
   \end{tabular}
 \end{table}

%


\begin{figure*}
  \centering
  \includegraphics[width=\linewidth]{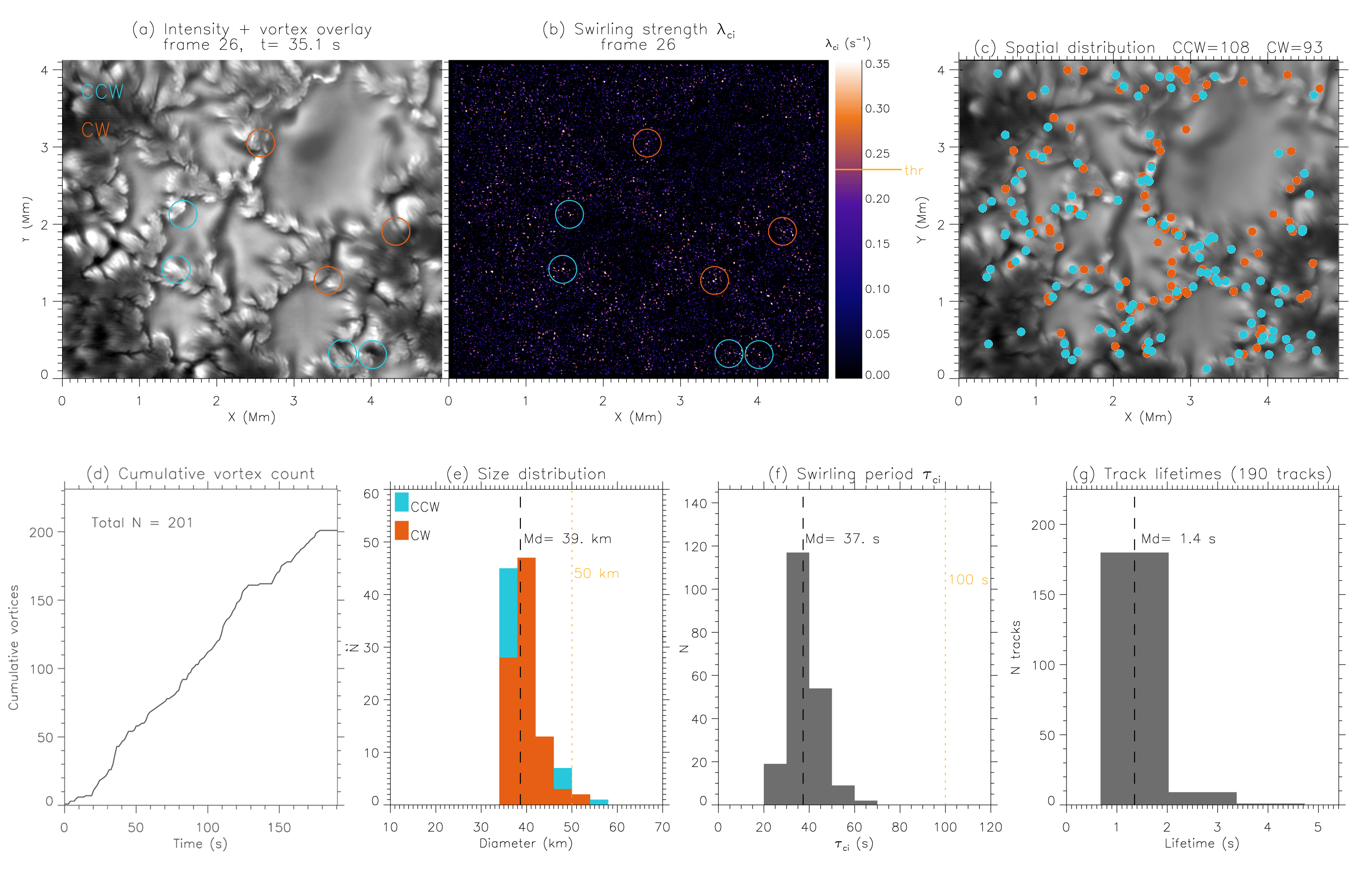}
  \caption{%
    Summary of the vortex detection pipeline results on the \dkist{} 416\,nm
    time series (143 frames, 3.2\,min, FOV = $4.92\times4.13$\,Mm).
    \textit{(a)}\,Intensity image at frame\,26 ($t = 35.1$\,s, peak count $N=7$)
    with detected vortices overlaid as circles
    (cyan: \ccw; orange: \cw).
    \textit{(b)}\,Swirling-strength map $\lci$ for the same frame. The yellow line
    in the color bar marks the detection threshold ($\lci^{\rm thr} =
    0.232$\,s$^{-1}$, 99th percentile); vortex circles follow the same colour
    convention as panel\,(a).
    \textit{(c)}\,Spatial distribution of all 201 vortex centroids accumulated
    over the full sequence, overlaid on the time-averaged intensity image (CCW: cyan; CW: orange).
    \textit{(d)}\,Number of detections per frame versus time. 
    \textit{(e)}\,Equivalent diameter distribution separated by rotation sense
    (\ccw: cyan; \cw: orange). Black dashed line: median diameter
    ($\tilde{d} = 38.6$\,km); orange dotted line: 50\,km lower bound of
    sub-granular vortices in \muramcode{} simulations \citep{Yadav2021}.
    \textit{(f)}\,Distribution of swirling periods $\tci = 2\pi/\lci$. Black
    dashed line: median ($\tilde{\tau}_{\rm ci} = 37.3$\,s); orange dotted
    line: $\tci = 100$\,s criterion of \citet{Yadav2021}.
    \textit{(g)}\,Track lifetime distribution for 190 resolved vortex tracks
    (black dashed: median 1.35\,s; maximum: 4.05\,s).}
  \label{fig:summary}
\end{figure*}

\section{Results}
\label{sec:results}

The pipeline detects 201 vortex instances across 142 frame pairs, with a mean
rate of 1.42\,frame$^{-1}$ and a spatial density of 0.070\,Mm$^{-2}$ over the
20.3\,Mm$^2$ FOV. The peak simultaneous count is 7 vortices at
frame\,26 ($t = 35.1$\,s; Fig.\,\ref{fig:summary}, panel\,(a)). The detection rate reflects the
stringency of the 99th percentile threshold, which retains only the most
intensely rotating structures.

The median equivalent diameter is $38.6$\,km (mean $39.5 \pm 3.3$\,km; 
Fig.\,\ref{fig:summary}, \\ panel\,(e)). Virtually all 201 detections
lie below 50\,km, and can be considered in the sub-granular regime.
All detections fall well below the 150\,km resolution limit of SUNRISE/IMaX
\citep{YellesChaouche2020}, confirming a population inaccessible to all prior
solar telescopes.

The median swirling period is $\tci = 37.3$\,s (Fig.\,\ref{fig:summary},
panel\,(f)), well below the $\tci < 100$\,s criterion of \citet{Yadav2021},
confirming dynamic consistency with the simulated sub-granular vortex
population. Of the 201 detections, 108 (53.7\%) are \ccw{} and 93 (46.3\%) are \cw. A
two-sided binomial test gives $p = 0.29$, which confirms no statistically
significant asymmetry (Fig.\,\ref{fig:summary}, panel\,(c)). This result is
consistent with the dominance of convective driving over the Coriolis force
at sub-granular scales.

The centroid map accumulated over the full sequence
(Fig.\,\ref{fig:summary}, panel\,(c)) shows a tendency toward intergranular
lane regions, consistent with \citet{Moll2011, Yadav2021}. The temporal tracking
yields 190 resolved tracks with a median lifetime of 1.35\,s and a maximum
of 4.05\,s (Fig.\,\ref{fig:summary}, panel\,(g)).

\section{Discussion}
\label{sec:discussion}


The vortices detected in this work match key predictions of \citet{Yadav2021} for a plage
region. The median diameter of 38.6\,km lies just below the lower end of the
predicted 50--100\,km range, consistent with the stringent 99th percentile
threshold selecting compact and intense vortex cores. The median $\tci = 37.3$\,s
lies well within the $\tci < 100$\,s criterion of \citet{Yadav2021}. The
co-temporal \hmi{} magnetogram confirms a unipolar plage of negative polarity
with $-50$ to $-400$\,G and, therefore, can be directly compared to the 200\,G simulation of \citet{Yadav2021}. The detected density of 0.070\,Mm$^{-2}$ is approximately one order of magnitude
below the ${\sim}1$\,Mm$^{-2}$ predicted by \citet{Moll2011}. We attribute this
to the imposed percentile threshold, which suppresses sensitivity for the detection of faint vortices.  Compared with prior observations, the vortices reported here are one order of
magnitude smaller in diameter than those of \citet{Bonet2008} and
\citet{Balmaceda2010}, and nearly a factor of twenty smaller than those of
\citet{VargasDominguez2011}. This directly reflects the fourfold improvement
in resolution afforded by \dkist{}.

Unlike the larger ($\geq$150 km) vortices reported in previous studies, for which lifetimes of tens to hundreds of seconds have been measured, the structures identified here have median diameters of only 38.6\,km. A simple scaling estimate, lifetime $\approx r_{\rm eq}/vflow$, using $r_{\rm eq} \lesssim 18$\,km and the intergranular flow speeds of 1--5\,km\,s$^{-1}$ measured here (Fig.\,\ref{fig:flowmap}a), yields characteristic lifetimes of order 4–20 s, comparable to our measured values and to the cadence of the observations. Also, the swirling period is an instantaneous diagnostic of the local rotation rate and is not expected to coincide with the tracked lifetime (see Methods\,\ref{sec:methods:lci}). We therefore interpret the short tracked lifetimes not as an artifact of the detection, but as a natural consequence of resolving a smaller-scale, faster-evolving vortex population than has not been accessible to previous instrumentation. In addition, visual inspection of the full-cadence movie reveals larger, more slowly rotating structures with longer apparent lifetimes at intergranular junctions, which correspond to vortices with $\lambda_{ci}$ below our 99th-percentile threshold and are not included in the statistics presented here. A dedicated lower-threshold analysis targeting this larger-covered-area, longer-lived population is left for future work.

The time-averaged horizontal velocity field (Fig.\,\ref{fig:flowmap}a) reveals
the granular flow pattern with typical outflows of 1--2\,km\,s$^{-1}$ from
granule centers and faster flows ($> 3$\,km\,s$^{-1}$) concentrated along the
intergranular lanes where vortex activity is also enhanced. This is consistent
with the picture of vortices being generated at the intersection of diverging
granular flows \citep{Moll2011}. 
We note that the horizontal velocity field used in this work is derived from FLCT, and that inferred vortex properties can depend on the velocity-estimation method, detection criterion, spatial resolution, and magnetic configuration \citep[e.g.,][]{Turkay2026, KollPistarini2026}. In this context, the stringent 99th-percentile $\lambda_{ci}$ threshold adopted here preferentially selects the strongest, most compact, and smallest-scale swirling signatures in the FLCT-derived velocity field.
\section{Conclusions}
\label{sec:conclusions}

We have presented the first direct observational detection of sub-granular
vortices in the solar photosphere using 416\,nm \dkist{} imagery. Applying
Fourier \lct{} and the swirling strength criterion to a 3.2-min sequence of
143 \mfbd-reconstructed frames over a $4.92 \times 4.13$\,Mm plage FOV,
we detect 201 vortex instances. Our main conclusions are summarized as follows (see also Tab.\,\ref{tab:results}):

\begin{enumerate}
  \item The median equivalent diameter is $38.6$\,km (mean $39.5 \pm 3.3$\,km).
        Virtually all 201 detections lie below 50\,km,
        and all lie well below the 150\,km resolution limit of SUNRISE, placing them within the 20–55 km size range predicted by MURaM simulations 
        \citep{Moll2011, Yadav2021} and previously inaccessible to direct observation.
  \item The median swirling period $\tci=37.3$\,s satisfies the $\tci < 100$\,s
        criterion of \muramcode{} simulations \citep{Yadav2021}.
  \item The spatial density of 0.070\,Mm$^{-2}$ is a conservative lower bound
        set by the 99th percentile threshold.
  \item We found no statistically significant rotation-sense asymmetry 
        (\ccw: 53.7\%, \cw: 46.3\%, $p = 0.29$), consistent with the
        dominance of convection over the Coriolis force at these scales.
  \item Temporal tracking yields 190 resolved tracks with a median lifetime of 1.35\,s and a maximum of 4.05\,s; given the observational cadence, these values represent lower bounds on the true vortex lifetimes (see Section\,\ref{sec:discussion}).
  \item The time-averaged horizontal velocity field resolves the granular
        convective pattern with typical flows of $1$--$5$\,km\,s$^{-1}$,
        providing direct evidence that the detected vortices are embedded in
        the intergranular convective network.
  \item These results demonstrate the unique resolving power of \dkist{} for
        solar convective dynamics and open a direct observational window on
        the sub-granular vortex population long predicted by
        magnetoconvection simulations.
\end{enumerate}

Future work will search for chromospheric counterparts of the detected 
vortices in the co-spatial \halpha{} 656\,nm and \caii{}\,IR observations. 
\muramcode{} simulations predict that photospheric vortices of this scale 
drive torsional Alfvén waves that produce observable brightenings and 
swirling motions in chromospheric diagnostics \citep{Yadav2021}, providing 
a direct observational test of the vortex-driven heating mechanism inferred 
from numerical models.

\begin{acknowledgements}
The authors thank the \dkist{} operations team at the National Solar Observatory
for support during the observations. The data were obtained as part of a
diagnostic test conducted in collaboration between NSO and MPS, G{\"o}ttingen.
\dkist{} is a facility of the National Science Foundation operated by NSO under
a cooperative agreement with AURA, Inc. The \hmi{} data are courtesy of
NASA/SDO and the HMI science team. The research was sponsored by the DynaSun project and has thus received funding under the Horizon Europe programme of the European Union under grant agreement (no. 101131534). Views and opinions expressed are however those of the author(s) only and do not necessarily reflect those of the European Union and therefore the European Union cannot be held responsible for them."

\end{acknowledgements}


\bibliographystyle{spr-mp-sola}
\bibliography{references}

\end{document}